\documentclass[10pt]{article}
\usepackage{amsmath}
\usepackage{graphicx}
\usepackage{color}
\usepackage{orcidlink}
\usepackage{hyperref}
\hypersetup{colorlinks=true, linkcolor=blue, citecolor=blue, urlcolor=blue}
\usepackage{subcaption}
\begin{document}
\baselineskip=14 pt

\begin{center}
{\large{\bf Quantum dynamics of spin-0 particles in a cosmological space-time }}
\end{center}

\vspace{0.3cm}

\begin{center}
    {\bf Faizuddin Ahmed\orcidlink{0000-0003-2196-9622}}\footnote{\bf faizuddinahmed15@gmail.com ; faizuddin@ustm.ac.in}\\
    \vspace{0.1cm}
    {\it Department of Physics, University of Science \& Technology Meghalaya, Ri-Bhoi, 793101, India}\\
    \vspace{0.2cm}
    {\bf Abdelmalek Bouzenada\orcidlink{0000-0002-3363-980X}}\footnote{\bf abdelmalek.bouzenada@univ-tebessa.dz ; abdelmalekbouzenada@gmail.com}\\
    \vspace{0.1cm}
    {\it  Laboratory of theoretical and applied Physics, Echahid Cheikh Larbi Tebessi University, Algeria}\\
\end{center}

\vspace{0.3cm}

\begin{abstract}
In this paper, our focus is on investigating the impact of cosmological constant on relativistic quantum systems comprising spin-0 scalar particles. Our analysis centers around the Klein-Gordon equation, and we obtain both approximate and exact analytical solutions for spin-0 particles of the quantum system. Afterwards, we explore quantum oscillator fields by considering the Klein-Gordon oscillator within the same space-time characterized by a cosmological constant. We obtain an approximate expression for the energy eigenvalue of the oscillator fields. In fact, the energy spectrum in both scenarios are examined and show the influences of the cosmological constant and geometry's topology. Our investigation is situated within the context of a magnetic universe-a four-dimensional cosmological space-time recognized as the Bonnor-Melvin universe. 
\end{abstract}

\vspace{0.1cm}


\vspace{0.1cm}


\section{Introduction}
\label{intro}

In modern physics, Einstein's revolutionary general theory of relativity (GR) skillfully paints gravity as an intrinsic geometric feature of space-time \cite{k1}. This conceptual framework unravels the mesmerizing correlation between space-time curvature and the genesis of classical gravitational fields, furnishing accurate forecasts for phenomena such as gravitational waves \cite{k2}, gravitational lensing \cite{k11,k12}, the Lense-Thirring effect exploring frame dragging \cite{k8,k9}, and black holes physics \cite{k3}. This theory deals with macroscopic objects or celestial objects. On the other hand, there is another branch of physics which deals with behaviors of particles at the microscopic scale known as quantum mechanics (QM) \cite{k4}. The convergence of these two domains of physics (general relativity and quantum mechanics) opens a gateway to profound insights into the fundamental nature of the universe. The triumph of quantum field theory in deciphering subatomic particle interactions and unraveling the origins of weak, strong, and electromagnetic forces \cite{k5} amplifies the anticipation surrounding this interdisciplinary fusion. Yet, the enduring quest for a unified theory-a theory of quantum gravity harmonizing general relativity and quantum mechanics-faces persistent challenges and technical intricacies, with recent strides \cite{k6,k7}. These challenges propel intense scientific endeavors as researchers diligently work to bridge remaining gaps, aspiring to unveil the foundational framework that unifies these two pivotal pillars of modern physics.

In recent times, there has been a noteworthy interest surrounding the investigation of magnetic fields, propelled by the discovery of systems boasting exceptionally potent fields, exemplified by magnetars \cite{k13,k14}, and occurrences observed in heavy ion collisions \cite{k15,k16,k17}. Exploring the integration of the magnetic field within the framework of general relativity raises intriguing questions. Presently, diverse solutions to the Einstein-Maxwell equations exist, including the Manko solution \cite{k18,k19}, the Bonnor-Melvin universe \cite{k20, k21}, the Bonnor-Melvin-likesolutions with cosmological constant \cite{MZ2, MZ3}, and a recently proposed electrovacuum solution \cite{MZ} that incorporates the cosmological constant into the Bonnor-Melvin framework. 

We are shifting our focus to the intersection of general relativity and quantum physics, a critical consideration emerges: how these two theories may interrelate or if such a connection is even relevant. Numerous works have addressed this inquiry, predominantly relying on the Klein-Gordon and the Dirac equations within curved space-times \cite{k23, k24}. This exploration extends to diverse scenarios, encompassing particles in Schwarzschild \cite{k25} and Kerr black holes \cite{k26}, cosmic string backgrounds \cite{k27,k28,k29}, quantum oscillators \cite{k30, k31, k32, k33, k34, k35, k35-1, k35-2, k35-3}, in the context of the Casimir effects \cite{k36, k37}, and particles within the Hartle-Thorne space-time \cite{k38}. Moreover, a few other investigations of quantum mechanical problems within the curved space-time backgrounds are in global monopoles \cite{hh, hh1, hh2, hh0}, in cosmic string space-time \cite{hh3}, in the Som-Raychaudhuri space-time \cite{hh4}, and other topological defect backgrounds \cite{hh5, hh6, hh7, pp1, pp2, pp3, pp4, pp5, pp6, pp7, pp8, pp9, pp10, pp11, pp12, pp13, pp14, k39, k40, k41}. These investigations have yielded compelling insights into how quantum systems respond to the arbitrary geometries of spacetime. Consequently, an intriguing avenue of study involves the examination of quantum particles within a space-time influenced by a magnetic field. For instance, in \cite{k42}, the Dirac particles were explored in the Melvin metric, while our current work focus into the study of dynamics of spin-0 bosonic fields within a magnetic universe incorporating a cosmological constant, as proposed in \cite{MZ}.

The dynamics of various relativistic quantum systems hinge critically on the Dirac oscillator (DO). As highlighted by Ito {\it et al.} \cite{k43}  in previous advancements in the dynamics of spin-1/2 particles with a linear trajectory, they demonstrated that the non-relativistic limit of the system results in an ordinary harmonic oscillator with a significant spin-orbit coupling term. Notably, Moshinsky and Szczepaniak \cite{k44} determined that the mentioned DO could be derived from the free Dirac equation by introducing an external linear potential, achieved through a minimal replacement of the momentum operator $\hat{p}\longrightarrow (\hat{p}-i\,M\,\omega\,\beta\,\hat{r})$. It is important to emphasize that, in addition to the theoretical exploration of the DO, valuable insights can be gained by delving into its physical interpretation-an aspect undeniably crucial for comprehending many relevant applications.

The Melvin magnetic universe is an exact static solution of Einstein-Maxwell equations in which Maxwell’s magnetic pressure is balanced by gravitational attraction. In this cylindrical symmetric space-time, the magnetic field is parallel to the regular axis of symmetry. The stability as well as other essential properties of Melvin’s magnetic universe have been extensively studied by a number of authors. The space-time metric in cylindrical coordinates is $ds^2=\Lambda^2 (r)\Big(-dt^2+dr^2+dz^2\Big)+\frac{r^2}{\Lambda^2(r)}\,d\phi^2$, where $\Lambda(r)=\Big(1+K^2\,r^2\Big)$ with $K=\frac{B}{2}$ and $B$ is the magnetic field strength. In the limit of zero magnetic field, $B \to 0$, one will get from this Bonnor-Melvin metric the Minkowski space given by $ds^2=-dt^2+dr^2+dz^2+r^2\,d\phi^2$.

The homogeneous solution of the Bonnor-Melvin-like space-time is described by the following line element (see Eq. (32) in Ref. \cite{MZ}) 
\begin{equation}
ds^{2}=g_{\mu\nu}dx^{\mu}dx^{\nu}=-dt^{2}+dz^2+\frac{1}{2\,\Lambda}\,\Big(dr^2+\alpha^2\,\sin^{2} r\,d\phi^{2}\Big),\label{a3}
\end{equation}
where $\Lambda$ denotes the cosmological constant, and $ 0 < \alpha <1$ represents the topological parameter which produces an angular deficit. Noted that the topological parameter $\alpha$ is introduced by modifying the angular coordinate $\phi$ via transformation $\phi \to \alpha\,\phi$. We assume a purely magnetic field aligned with the axis of symmetry, yielding thus a Maxwell tensor of the form $F=H(r)\,dr \wedge d\phi$, where the magnetic field strength is given by (see Eq. (11) in Ref. \cite{MZ})
\begin{equation}
H (r)=\frac{\alpha}{\sqrt{2}}\,\sin r\,.\label{a4}
\end{equation}
The covariant ($g_{\mu\nu}$) and contravariant form ($g_{\mu\nu}$) of the metric tensor for the space-time (\ref{a3}) are given by
\begin{eqnarray}
g_{\mu\nu}=\begin{pmatrix}
-1 & 0 & 0 & 0\\
0 & \frac{1}{2\,\Lambda} & 0 & 0\\
0 & 0 & \frac{\alpha^2\,\sin^2 r}{2\,\Lambda} & 0\\
0 & 0 & 0 & 1
\end{pmatrix},\quad 
g^{\mu\nu}=\begin{pmatrix}
-1 & 0 & 0 & 0\\
0 & 2\,\Lambda & 0 & 0\\
0 & 0 & \frac{2\,\Lambda}{\alpha^2\,\sin^2 r} & 0\\
0 & 0 & 0 & 1
\end{pmatrix}. \label{a5}
\end{eqnarray}
The determinant of the metric tensor for the space-time (\ref{a4}) is given by 
\begin{equation}
    det\,(g_{\mu\nu})=g=-\frac{\alpha^2}{4\,\Lambda^2}\,\sin^2 r\,. \label{a6}
\end{equation}

The investigation into gravitational interactions within a quantum mechanical system has been an active area of research, as evidenced by numerous studies \cite{aa1,aa2,aa3,aa4,aa5}. The exploration of quantum mechanical phenomena in curved space-time introduces a novel interplay between quantum matter and gravitational forces in the micro-particle realm. Notably, recent research has been inspired by the Dirac oscillator embedded in a cosmic string background, leading to a wealth of studies such as the dynamics of the Dirac oscillator in cosmic string space-time \cite{aa8,aa9,aa3,k29,aa12,aa13}, the Aharonov-Casher effect on the Dirac oscillator \cite{aa1,bb1}, and the examination of non-inertial effects on the Dirac oscillator in the backdrop of cosmic string space-time \cite{bb2,bb3}.

Our motivation is to investigate the relativistic quantum dynamics of scalar particles within the context of a magnetic universe, specifically described by the Bonnor-Melvin metric with a cosmological constant. We aim to solve the Klein-Gordon wave equation analytically and obtain the exact and approximate eigenvalue solutions for the scalar particles. Subsequently, we explore the relativistic quantum oscillator, modeled by the Klein-Gordon oscillator, within the same magnetic universe background. We derive the radial wave equation and obtain the approximate eigenvalue solution for the oscillator field and conduct a comprehensive analysis of the results. Our findings reveal that the energy eigenvalues in both analyses are influenced by the topology of the geometry, as characterized by the parameter $\alpha$, and the cosmological constant $\Lambda$. Furthermore, in the quantum oscillator system, the energy spectrum undergoes modifications due to the oscillator frequency $\omega$, in addition to the aforementioned parameters. The structure of this paper is summarized as: In {\tt Sect. 2}, we investigate the relativistic quantum dynamics of scalar particle via the Klein-Gordon equation in the background of magnetic universe. In {\tt Sect. 2}, we focus on the relativistic quantum oscillator in the background of same magnetic universe. In {\tt Sect. 4}, we present our results and discussion. Throughout this paper, we use the system of units, where $\hbar=c=G=1$.

\section{Dynamics of scalar particles in cosmological space-time background }

In this part, we study the relativistic quantum dynamics of spin-0 scalar particles described the Klein-Gordon wave equation in the background of cosmological space-time represented by the line-element (\ref{a3}). We derive that radial equation through a wave function ansatz and obtain approximate and exact eigenvalue solutions of the quantum system under investigations.

Therefore, we write the relativistic wave equation describing the quantum motions of spin-0 scalar particles given by \cite{k27, k28, k29, WG}
\begin{eqnarray}
    \Bigg[-\frac{1}{\sqrt{-g}}\,\partial_{\mu}\,\Big(\sqrt{-g}\,g^{\mu\nu}\,\partial_{\nu}\Big)+M^2\Bigg]\,\Psi=0,\label{b1}
\end{eqnarray}
where $M$ is the rest mass of the particles, $g$ is the determinant of the metric tensor $g_{\mu\nu}$ with its inverse $g^{\mu\nu}$. 

Expression the wave equation (\ref{b1}) in the space-time (\ref{a3}) and using Eqs. (\ref{a5})-(\ref{a6}), we obtain the following differential equation:
\begin{eqnarray}
    \Bigg[-\frac{d^2}{dt^2}+2\,\Lambda\,\Bigg\{\frac{d^2}{dr^2}+\frac{1}{\tan r}\,\frac{d}{dr}+\frac{1}{\alpha^2\,\sin^2 r}\,\frac{d^2}{d\phi^2}\Bigg\}+\frac{d^2}{dz^2}-M^2\Bigg]\,\Psi (t, r, \phi, z)=0\,.\label{b2}
\end{eqnarray}

In quantum mechanical system, the total wave function is expressible in terms of different variables by the method of separation of variables. In our case, we choose the following ansatz of the total wave function $\Psi$ in terms of function $\psi (x)$ as follows:
\begin{equation}
    \Psi (t, r, \phi, z)=\exp(-i\,E\,t)\,\exp(i\,m\,\phi)\,\exp(i\,k\,z)\,\psi(r), \label{b3}
\end{equation}
where $E$ is the particle's energy, $m=0,\pm\,1,\pm\,2,....$ are the eigenvalues of the angular quantum number. We can express the total wave function in the given form due to the absence of dependence on the time coordinate $t$, the angular coordinate $\phi$, and the translational coordinate $z$ in the differential equation (\ref{b2}).

Substituting the total wave function (\ref{b3}) into the differential equation (\ref{b8}), we obtain the differential equation for $\psi(x)$ as follows:
\begin{equation}
    \psi'' (r)+\frac{1}{\tan r}\,\psi' (r)+\Bigg[\frac{1}{2\,\Lambda}\Big\{E^2-k^2-M^2\Big\}-\frac{m^2}{\alpha^2\,\sin^2 r}\Bigg]\,\psi (r)=0\,.\label{b4}
\end{equation}

Below, we endeavor to solve the aforementioned second-order differential equation (\ref{b4}) employing two distinct methods. The initial approach involves utilizing an approximation technique, while the second method employs an exact solution. We derive the expression for the energy eigenvalue using both methods and subsequently conduct a detailed analysis of the obtained results.

\subsection{Approximate Eigenvalue Solution}

In this section, we address the equation (\ref{b4}) by employing a first-order approximation, specifically $\sin r \approx r$ and $\tan r \approx r$. This approximation is applicable only under the condition that the radial distance $r$ is sufficiently small, confining the scalar particles within this region. Consequently, the radial wave equation (\ref{b4}) simplifies to the following form:
\begin{equation}
    r^2\,\psi''+r\,\psi'+\Big(\lambda^2\,r^2-\iota^2\Big)\,\psi=0,\label{b5}
\end{equation}
where we have defined
\begin{equation}
\lambda=\sqrt{\frac{1}{2\,\Lambda}\,(E^2-k^2-M^2)},\quad \iota=\frac{|m|}{|\alpha|}\,.\label{ss}    
\end{equation}

It is evident that the resulting differential equation (\ref{b5}) takes on the well-known form of Bessel's second-order equation \cite{MA,GBA}, for which solutions are readily available. Our primary focus lies in obtaining a well-behaved solution of the Bessel equation at the origin, i.e., when $r=0$. The Bessel function of the first kind emerges as a viable solution to equation (\ref{b5}), exhibiting regular behavior at the origin, and can be expressed as:
\begin{equation}
    \psi (r)=c_1\,J_{\iota} (\lambda\,r),\label{b6}
\end{equation}
where $c_1$ is an arbitrary constant. The asymptotic form of the Bessel function of the first kind is given by \cite{MA, GBA}:
\begin{equation}
    J_{\iota} \propto \cos \Big(\lambda\,r-\frac{\pi}{2}\,\iota-\frac{\pi}{4}\Big)\,.\label{b7}
\end{equation}

Now, we confine the motion of scalar particles within a region characterized by a hard-wall confining potential. This confinement is particularly significant as it provides an excellent approximation when investigating the quantum properties of systems such as gas molecules and other particles that are inherently constrained within a defined spatial domain. The hard-wall confinement is defined by a condition specifying that at a certain axial distance, $r=r_0$, the radial wave function $\psi$ becomes zero, i.e., $\psi (r=r_0)=0$. This condition is commonly referred to as the Dirichlet's condition in the literature. The study of the hard-wall confining potential has proven valuable in various contexts, including its examination in the presence of rotational effects on the scalar field \cite{k29}, studies involving the Klein-Gordon oscillator subjected to the influence of linear topological defects \cite{RLLV2, k28}, investigations into non-inertial effects on a non-relativistic Dirac particle \cite{bb2}, examinations of a Dirac neutral particle analogous to a quantum dot \cite{RLLV6}, studies on the harmonic oscillator within an elastic medium featuring a spiral dislocation \cite{RLLV7}, and investigations into the behavior of Dirac and Klein-Gordon oscillators in the presence of a global monopole \cite{hh}. This exploration of the hard-wall potential in diverse scenarios enriches our understanding of its impact on quantum systems, providing insights into the behavior of scalar particles subject to this form of confinement. 

Therefore, at $r=r_0$, we have $\psi(r=r_0)=0$ and using Eq. (\ref{b7}), we obtain the following relation:
\begin{equation}
    \Big(\lambda\,r_0-\frac{\pi}{2}\,\iota-\frac{\pi}{4}\Big)=(2\,n+1)\,\frac{\pi}{2},\label{b8}
\end{equation}
where $n=0,1,2,3,...$. 

Simplification of the above relation gives us the energy eigenvalue of scalar particles associated with the mode $\{n, m\}$ given by:
\begin{equation}
    E_{n,m}=\pm\,\sqrt{k^2+M^2+\frac{2\,\Lambda\,\pi^2}{r^2_{0}}\,\Big(n+\frac{|m|}{2\,|\alpha|}+\frac{3}{4}\Big)^2}\,.\label{b9}
\end{equation}

\begin{center}
\begin{figure}
\begin{centering}
\subfloat[$\alpha=0.5,m=1$]{\centering{}\includegraphics[scale=0.5]{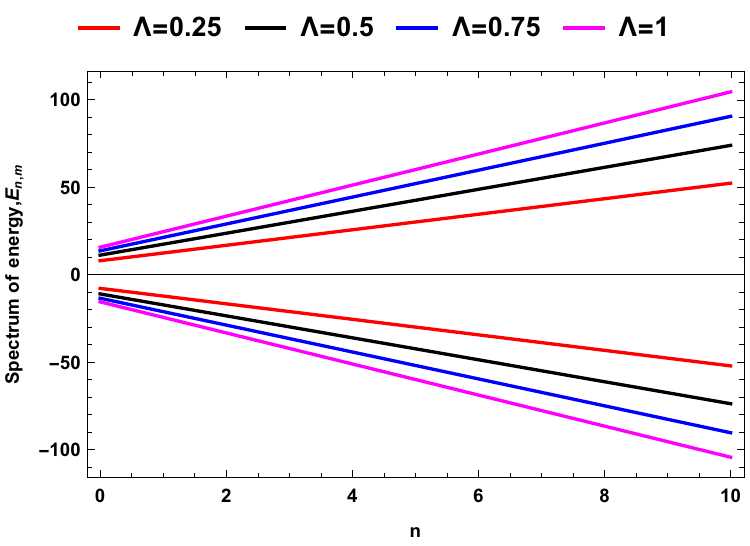}}\quad\quad
\subfloat[$\Lambda=0.5,m=1$]{\centering{}\includegraphics[scale=0.5]{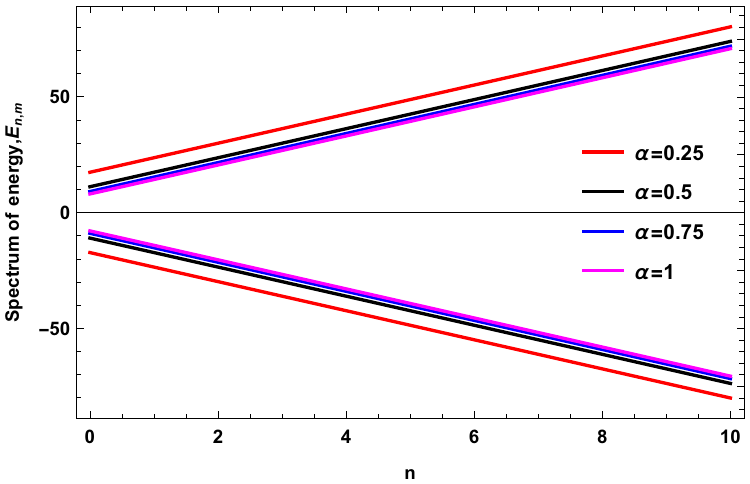}}
\par\end{centering}
\begin{centering}
\subfloat[$\varLambda=\alpha=0.5$]{\centering{}\includegraphics[scale=0.5]{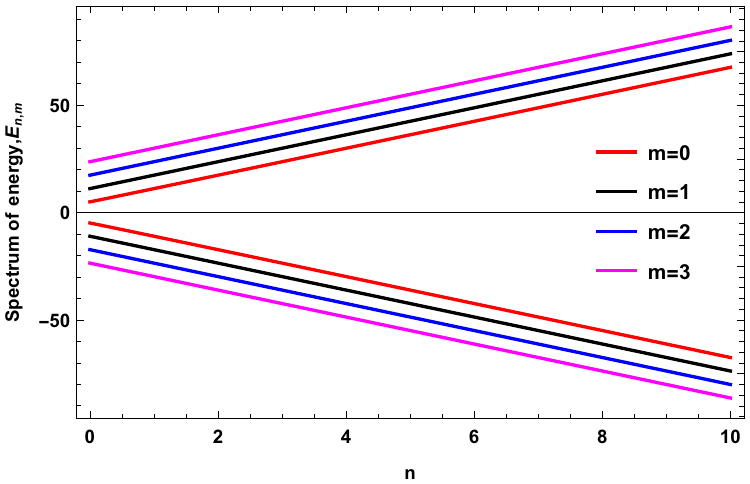}}\quad\quad
\subfloat[$\alpha=0.5,m=1$]{\centering{}\includegraphics[scale=0.5]{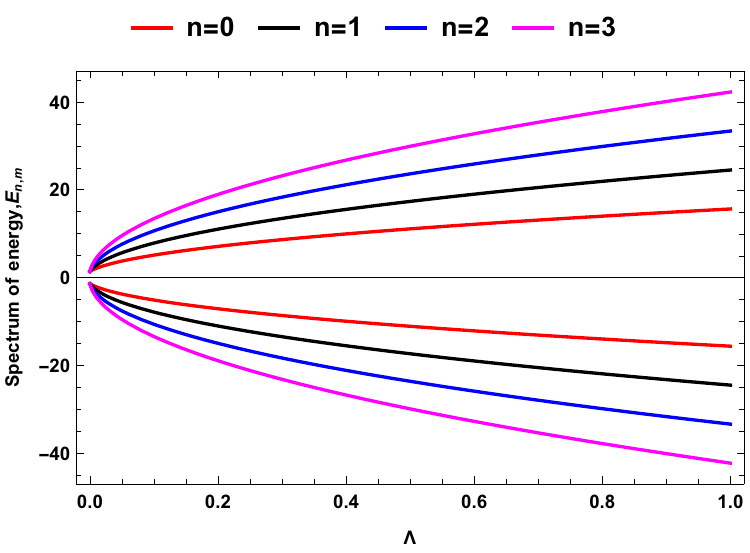}}
\par\end{centering}
\centering{}\caption{Energy spectrum $E_{m,n}$ (equation (\ref{b9})). Here the parameters are set as $k=M=1$ and $r_{0}=0.5$.}
\hfill\\
\begin{centering}
\subfloat[]{\centering{}\includegraphics[scale=0.5]{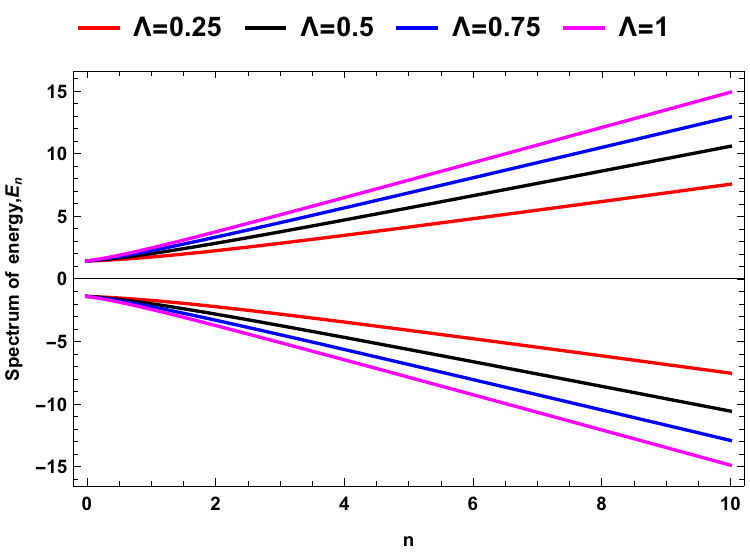}}\quad\quad
\subfloat[]{\centering{}\includegraphics[scale=0.5]{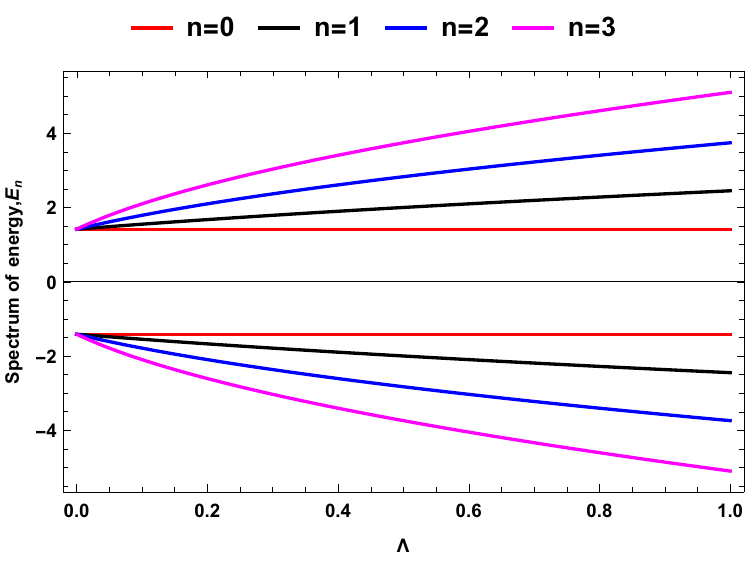}}
\par\end{centering}
\centering{}\caption{Energy spectrum $E_{n}$ (equation (\ref{b14})). Here the parameters are set as $k=M=1$.}
\end{figure}
\par\end{center}

Equation (\ref{b9}) represents the relativistic approximate energy eigenvalue of spin-0 scalar particles in the context of cosmological space-time described by the line-element (\ref{a3}). Notably, the approximate energy spectrum is intricately shaped by the underlying geometry, as characterized by the parameters $\alpha$ and $\Lambda$. Additionally, the energy spectrum exhibits variations with respect to the angular quantum number $m$. A noteworthy observation is that the energy levels of the scalar particles exhibit equal spacing on both sides of $E=0$. This intriguing feature signifies that in the system under consideration, particles and antiparticles possess equal energies. This symmetry in energy levels adds a distinctive characteristic to the behavior of scalar particles in this cosmological space-time, emphasizing the equilibrium between particles and their corresponding antiparticles within this specific framework.

We have generated Figure 1 to illustrate the energy spectrum for various values of the cosmological constant $\lambda$ (Fig. 1(a)), while keeping $\alpha=0.5$ and $m=1$. Additionally, Figure 1(b) depicts the impact of the topological parameter $\alpha$ on the energy spectrum, with $\Lambda=0.5$ and $m=1$. In Figure 1(c), we explore the influence of the angular quantum number $m$ on the energy spectrum, maintaining fixed values of $\alpha=0.5=\Lambda$. Lastly, Figure 1(d) illustrates the variation in the energy spectrum with changes in the radial quantum number $n$, with $\alpha=0.5$ and $m=1$. These figures showcase the nature of the energy spectrum as these parameters $(\Lambda, \alpha, m, n)$ increase, providing valuable insights into the behavior of the system under consideration.

\subsection{Exact Eigenvalue Solution}

In this part, we address the equation (\ref{b4}) by making a transformation $u=\cos r$ into the differential equation (\ref{b4}) results the following form:
\begin{equation}
    (1-u^2)\,\psi''(u)-2\,u\,\psi'(u)+\Big(\lambda-\frac{\iota^2}{1-u^2}\Big)\,\psi (u)=0\,.\label{b10}
\end{equation}
Setting $\ell=\Big(\lambda+\frac{1}{4}\Big)^{\frac{1}{2}}-\frac{1}{2}$ which results $\ell\,(\ell+1)=\lambda$ into the above equation (\ref{b10}), we obtain
\begin{equation}
    (1-u^2)\,\psi''(u)-2\,u\,\psi'(u)+\Big[\ell\,(\ell+1)-\frac{\iota^2}{1-u^2}\Big]\,\psi (u)=0\,.\label{b11}
\end{equation}
The above differential equation is the well-know associated Legendre polynomial \cite{MA,GBA} whose solutions are called associated Legendre polynomials given by:
\begin{equation}
    \psi(u)=P^{\iota}_{\ell} (u)=(-1)^{\iota}\,(1-u^2)^{\frac{\iota}{2}}\,\frac{d^{\iota}}{du^{\iota}}\,(P_{\ell} (u)),\label{b12}
\end{equation}
where $P_{\ell} (u)$ is the Legendre polynomial.

It is known that this associated Legendre polynomial can be expressed in terms of hypergeometric function ${}_2F_1 \Big(-\ell, \ell+1;1-\iota;\frac{1-u}{2}\Big)$ \cite{MA} and by series expansion one can show that this hypergeometric function ${}_2F_1$ becomes a finite degree polynomial of degree $n$ provided $-\ell$ must be a non-positive integer, that is $\ell=n$, where $n=0,1,2,3,...$. Thus, we can write $\ell\,(\ell+1)=\lambda=n\,(n+1)$.

Simplification of the above relation $\lambda=n\,(n+1)$ results the following expression of the energy eigenvalue of the system given by:
\begin{equation}
    E_{n}=\pm\,\sqrt{M^2+k^2+2\,\Lambda\,n\,(n+1)}\,.\label{b14}
\end{equation}

Equation (\ref{b14}) represents the exact relativistic energy eigenvalue of spin-0 scalar particles within the cosmological space-time described by (\ref{b1}). It is apparent that this exact energy expression is solely influenced by the cosmological constant $\Lambda$ and exhibits variations dependent on the radial quantum number $n$. Noteworthy is the observation that the energy levels of the scalar particles display equidistant spacing on both sides around $E=0$, indicating an equality in energies between particles and antiparticles within the considered system.

It is important to highlight the disparity between the approximate and exact energy eigenvalues of scalar particles, as expressed respectively in Eqs. (\ref{b9}) and (\ref{b14}). Furthermore, there exists a possibility of obtaining an approximate energy eigenvalue by solving equation (\ref{b4}) up to the second-order approximation, a consideration omitted in this paper. Future investigations may explore this avenue to provide a more comprehensive understanding of the system's behavior.

We have presented Figure 2 to depict the exact energy spectrum of scalar particles for different values of the cosmological constant $\lambda$ (Fig. 2(a)) and the radial quantum number $n$ (Fig. 2(b)). The figures illustrate the behavior of the exact energy spectrum as these parameters $(\Lambda, n)$ increase, offering insights into the system's dynamics with varying cosmological constants and radial quantum numbers. 

\section{Relativistic quantum oscillator fields in cosmological space-time background }

In this section, we study the relativistic quantum oscillator fields within the framework of the Klein-Gordon oscillator, considering the backdrop of the cosmological solution given by (\ref{a3}). To account for the oscillator field, we introduce a modification by replacing the radial momentum operator $\frac{\partial}{\partial r} \to \Big(\frac{\partial}{\partial r}+M\,\omega\,r\Big)$, where $\omega$ represents the oscillator frequency \cite{k30, k31, k32, k33, k34, k35, k35-1, k35-2, k35-3}. This adjustment is further expressed in four-vector form as $\partial_{\mu} \to (\partial_{\mu}+M\,\omega\,X_{\mu})$, with $X_{\mu}=(0, r, 0, 0)$. Numerous researchers have extensively explored the dynamics of this relativistic quantum oscillator field in various space-time backgrounds, including G\"{o}del and G\"{o}del-type space-times both with and without topological defects. Investigations have been conducted in the context of global monopoles space-time, cosmic string space-times (both standard and spinning), as well as in topologically trivial and non-trivial geometries. Additionally, studies within the realm of Kaluza-Klein theory (KKT) have been undertaken (see, Refs. \cite{k30, k31, k32, k33, k34, k35, k35-1, k35-2, k35-3, hh, hh1, hh2, hh0}, and the associated references therein). The examination of the relativistic quantum oscillator field within diverse space-time backgrounds provides valuable insights into the interplay between quantum oscillation fields and the underlying cosmological geometry, offering a comprehensive understanding of the intricate dynamics involved in these scenarios.

The relativistic wave equation describing the quantum oscillator field is given by 
\begin{eqnarray}
    \Big[\frac{1}{\sqrt{-g}}\,(\partial_{\mu}+M\,\omega\,X_{\mu})\,(\sqrt{-g}\,g^{\mu\nu})\,(\partial_{\nu}-M\,\omega\,X_{\nu})\Big]\,\Psi=M^2\,\Psi,
    \label{c1}
\end{eqnarray}
where $\omega$ is the oscillator frequency. 

Explicitly writing the wave equation (\ref{c1}) in the space-time (\ref{a3}) and using Eqs. (\ref{a5})-(\ref{a6}), we obtain the following differential equation:
\begin{eqnarray}
    \Bigg[-\frac{d^2}{dt^2}+2\,\Lambda\,\Bigg\{\frac{d^2}{dr^2}+\frac{1}{\tan r}\,\frac{d}{dr}-M\,\omega-\frac{M\,\omega\,r}{\tan r}-M^2\,\omega^2\,r^2+\frac{1}{\alpha^2\,\sin^2 r}\,\frac{d^2}{d\phi^2}\Bigg\}+\frac{d^2}{dz^2}-M^2\Bigg]\,\Psi=0\,.\label{c2}
\end{eqnarray}
Substituting the wave function ansatz (\ref{b3}) into the above differential equation (\ref{c2}) results the following second-order differential equation form:
\begin{equation}
    \psi''+\frac{1}{\tan r}\,\psi'+\Bigg[\frac{(E^2-M^2)}{2\,\Lambda}-M\,\omega-\frac{M\,\omega\,r}{\tan r}-M^2\,\omega^2\,r^2-\frac{k^2}{2\,\Lambda}-\frac{\iota^2}{\sin^2 r}\Bigg]\,\psi=0.\label{c3}
\end{equation}

\begin{center}
\begin{figure}
\begin{centering}
\subfloat[$\alpha=0.5,m=1$ ]{\centering{}\includegraphics[scale=0.5]{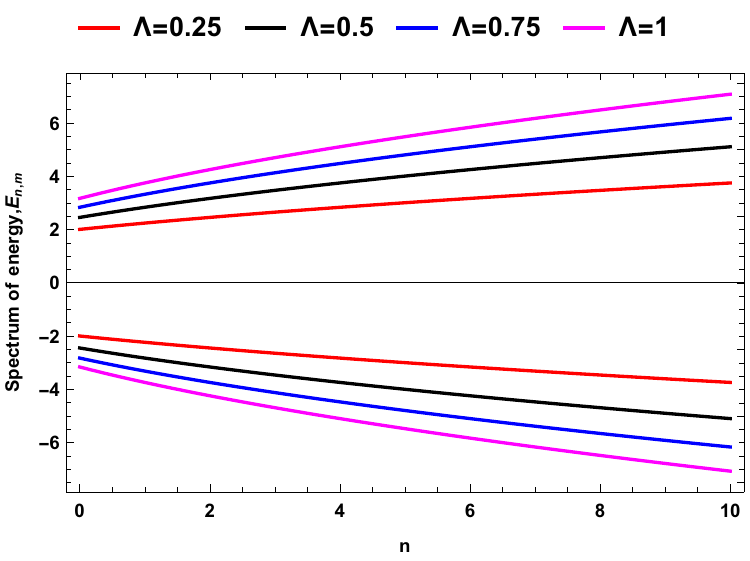}}\quad\quad
\subfloat[$\varLambda=0.5,m=1$]{\centering{}\includegraphics[scale=0.5]{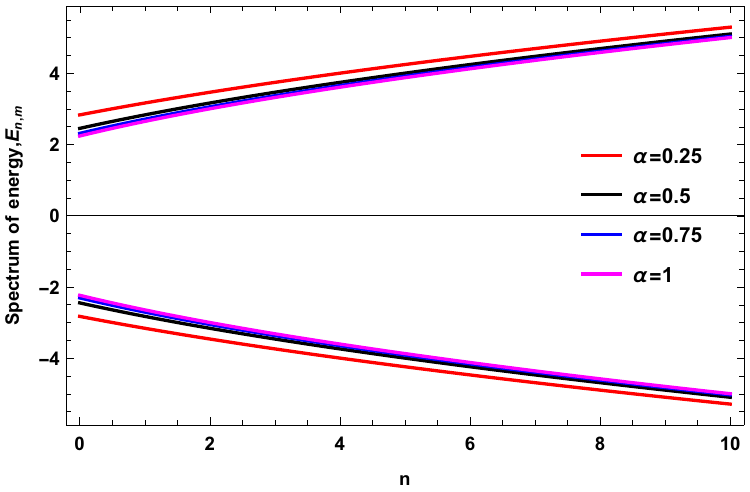}}
\par\end{centering}
\begin{centering}
\subfloat[$\varLambda=\alpha=0.5$]{\centering{}\includegraphics[scale=0.5]{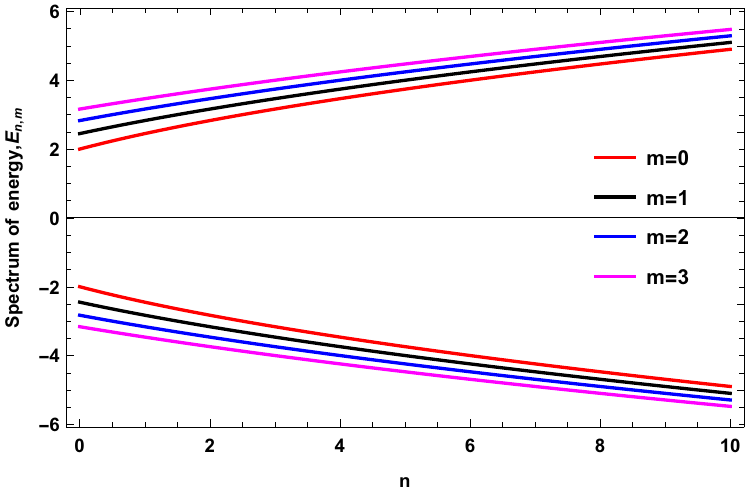}}\quad\quad
\subfloat[$n=0,\alpha=0.5,m=1$]{\centering{}\includegraphics[scale=0.5]{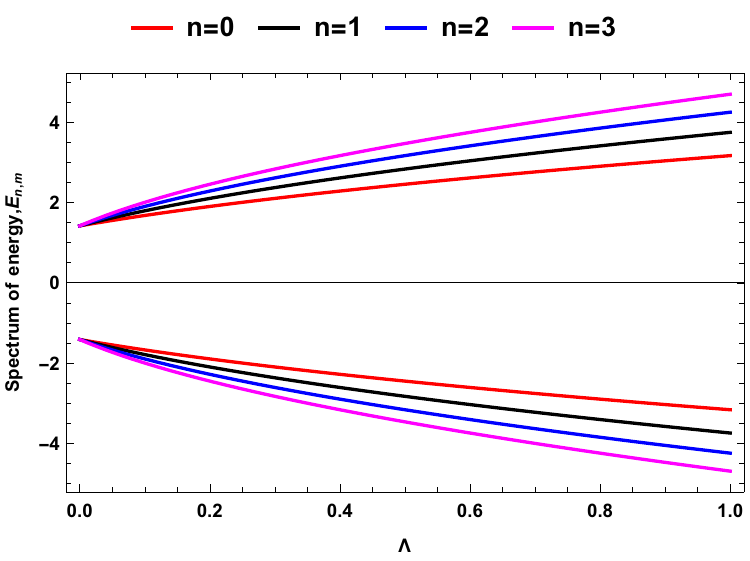}}
\par\end{centering}
\centering{}\caption{The energy spectrum $E_{n,m}$ for the relation given in Equation \ref{c7}, where the parameters are set as $k=M=1$ and $\omega=0.5$.}
\end{figure}
\par\end{center}

We solve the above differential equation (\ref{c3}) by taking an approximation up to the first order, that is, $\sin r \approx r$ and $\tan r \approx r$. Therefore, the radial wave equation (\ref{c3}) reduces to the following form:
\begin{equation}
    \psi''+\frac{1}{r}\,\psi'+\Bigg(\eta^2-M^2\,\omega^2\,r^2-\frac{\iota^2}{r^2}\Bigg)\,\psi=0,\label{c4}
\end{equation}
where 
\begin{equation}
\eta^2=\frac{E^2-M^2-k^2}{2\,\Lambda}-2\,M\,\omega\,.\label{kk}    
\end{equation}

Transforming to a new variable via $s=M\,\omega\,r^2$ into the above equation (\ref{c4}) results the following equation form \cite{AFN}:
\begin{equation}
\psi''(s)+\frac{(c_1-\,c_2\,s)}{s\,(1-c_3\,s)}\,\psi'(s)+\frac{\Big(-\xi_1\,s^2+\xi_2\,s-\xi_3\Big)}{s^2\,(1-c_3\,s)^2}\,\psi (s)=0,\label{c5}
\end{equation}
where $c_1=1$, $c_2=0=c_3$ and 
\begin{equation}
\xi_1=\frac{1}{4},\quad \xi_2=\frac{\eta}{4\,M\,\omega},\quad \xi_3=\frac{\iota^2}{4}.\label{c6}
\end{equation}

Equation (\ref{c5}) takes the form of a homogeneous second-order differential equation, and we can solve it using the parametric Nikiforov-Uvarov method \cite{AFN}. The parametric Nikiforov-Uvarov method has been widely employed by various authors to obtain eigenvalue solutions for wave equations in quantum systems (see, for example, Refs. \cite{hh1, hh2}). Following the methodology outlined in Refs. \cite{hh1, hh2}, we successfully derive the relativistic approximate energy eigenvalue for the quantum system, and it is expressed as
\begin{equation}
   E_{n,m}=\sqrt{M^2+k^2+8\,M\,\omega\,\Lambda\,\Big(n+1+\frac{|m|}{2|\alpha|}\Big)},\label{c7}
\end{equation}
where $n=0,1,2,....$.

The corresponding radial wave function will be
\begin{equation}
    \psi(s)=\mathcal{N}\,s^{\frac{|m|}{2|\alpha|}}\,e^{-s/2}\,L^{(\frac{|m|}{|\alpha|})}_{n} (s),\label{c8}
\end{equation}
where $\mathcal{N}$ is the normalization constant.

Equation (\ref{c7}) characterizes the relativistic approximate energy eigenvalue of the quantum oscillator fields within the cosmological space-time background with topological defect. Notably, the expression for energy spectrum is influenced by the underlying geometry, as described by the topology parameter $\alpha$, and the cosmological constant $\Lambda$. In addition, the energy spectrum of the oscillator fields undergoes modifications introduced by the oscillator frequency $\omega$ and reveals variations in relation to the quantum numbers $\{n, m\}$. Here also, we see that the energy levels are equally spacing on both sides about $E=0$. 

We have presented Figure 3, illustrating the energy spectrum for different values of the cosmological constant $\Lambda$ (Fig. 3(a)), where $\alpha=0.5$ and $m=1$. Additionally, Figure 3(b) demonstrates the impact of the topological parameter $\alpha$ on the energy spectrum, with $\Lambda=0.5$ and $m=1$. Figure 3(c) portrays the influence of the angular quantum number $m$ on the energy spectrum while maintaining fixed values of $\alpha=0.5$ and $\Lambda=0.5$. Lastly, Figure 3(d) illustrates the variation in the energy spectrum with changes in the radial quantum number $n$, while $\alpha=0.5$ and $m=1$. The figures reveal that the linear nature of the energy spectrum undergoes changes with increasing values of these parameters $(\Lambda, \alpha, m, n)$, providing valuable insights into the evolving dynamics of the oscillator fields under consideration.

\section{Conclusions}

Numerous investigations have been conducted to explore exact solutions to the Einstein-Maxwell field equations, both in the presence and absence of a cosmological constant, spanning various dimensions. These solutions typically emerge from the electromagnetic field components aligned along axial and azimuthal directions. The pioneering work of Melvin marked the inception of solutions featuring a purely magnetic field, a concept later generalized into the Bonnor-Melvin magnetic universe. Notably, recent contributions in Refs. \cite{MZ2, MZ3, MZ} expanded upon this framework by introducing a non-zero cosmological constant into the Bonnor-Melvin magnetic universe-a topic of particular interest within the context of quantum mechanical systems. The inclusion of a cosmological constant in the Bonnor-Melvin magnetic universe opens avenues for investigating quantum mechanical phenomena within this unique spacetime. Such studies offer valuable insights into the dynamics of scalar particles and relativistic quantum oscillator fields, elucidating their behavior under the combined influences of the cosmological constant and the topology parameter. This exploration provides a nuanced understanding of the intricate interplay between quantum effects and the underlying geometry of the magnetic universe.

The primary objective of this study was to probe the behavior of spin-0 scalar particles and quantum oscillator fields within the context of a cosmological space-time characterized by the line-element (\ref{a3}), incorporating a non-zero cosmological constant. The central emphasis was placed on unraveling the impacts on the total wave function-a crucial entity that encapsulates comprehensive information about the quantum system under scrutiny. By delving into this investigation, our aim was to gain insights into the nuanced dynamics of scalar particles and quantum oscillator fields within the distinctive backdrop of this cosmological space-time. The focal point of interest revolved around understanding how the presence of a non-zero cosmological constant influences the total wave function, thereby providing a deeper comprehension of quantum mechanical phenomena within the framework of this magnetic universe. Through this exploration, our study sought to contribute to the broader understanding of the intricate interplay between the unique features of the cosmological space-time, characterized by the line-element (\ref{a3}), and the behavior of quantum systems. By shedding light on the behavior of scalar particles and quantum oscillator fields in this context, we aimed to enhance our comprehension of the underlying quantum mechanical phenomena within such specialized space-time configurations.

In Section 2, we successfully derived the radial equation of the Klein-Gordon equation within the Bonnor-Melvin magnetic universe featuring a cosmological constant. Subsequently, we solved this radial equation, presenting both an approximate energy eigenvalue, as outlined in Equation (\ref{b9}), and an exact analytical solution given by Equation (\ref{b14}). Notably, our observations revealed that the energy spectrum presented in Equation (\ref{b9}) is notably influenced by both the topology of the geometry ($\alpha$) and the presence of a positive cosmological constant ($\Lambda > 0$). Furthermore, this energy spectrum undergoes modifications based on the angular quantum number ($m$). An interesting contrast arises when comparing this energy spectrum (Equation (\ref{b9})) with the one presented in Equation (\ref{b4}). The latter is intriguingly independent of both the topological parameter ($\alpha$) and the angular quantum number ($m$). This discrepancy adds a layer of complexity to the understanding of how the geometry and cosmological constant influence the energy levels of the quantum system. Of particular note is the fascinating observation that the energy levels, as described in Equation (\ref{b9}), exhibit equal spacing on either side of $E=0$. This intriguing pattern suggests that particles and their corresponding antiparticles possess equivalent energies within fixed quantum states $\{n, m\}$. This finding adds a unique dimension to our understanding of the energy distribution within the quantum system, emphasizing the symmetry between particles and antiparticles for specific quantum states.

In Section 3, our investigation extended to the realm of relativistic quantum oscillator fields, specifically described by the Klein-Gordon oscillator. Within this framework, we derived the radial equation and successfully obtained the approximate energy eigenvalue for the oscillator fields, as detailed in Equation (\ref{c7}). Similar to our earlier findings, we noted that the energy spectrum is not only influenced by the topology of the geometry ($\alpha$) but is also impacted by the presence of a positive cosmological constant ($\Lambda > 0$) and the oscillator frequency ($\omega$). Additionally, the energy spectrum experiences alterations based on the angular quantum number ($m$). An intriguing parallel emerged when comparing the energy spectrum in Equation (\ref{c7}) with the one discussed in Section 2 (Equation (\ref{b9})). Once again, we observed the noteworthy pattern of energy levels exhibiting equal spacing on both sides of $E=0$. This consistent observation suggests that, akin to the scalar particles discussed earlier, particles and antiparticles within the relativistic quantum oscillator fields possess equivalent energies for fixed quantum states $\{n, m\}$. This recurring symmetry in the energy distribution underscores a fascinating aspect of the quantum dynamics within this cosmological framework.

Our exploration aimed into the intriguing realm of quantum field theory in curved space, with a specific focus on a cosmological space-time characterized by a non-zero positive cosmological constant. Throughout our investigation, we discerned that the outcomes exhibit modifications influenced by various parameters inherent in the space-time geometry. The interplay of these parameters, such as the topology parameter $\alpha$, cosmological constant $\Lambda$, and quantum numbers $\{n, m\}$, introduced nuanced changes to the results, adding layers of complexity to the quantum dynamics within this cosmological setting.


\section*{Conflict of Interest}

There is no conflict of interests.

\section*{Funding Statement}

No funding agency is associated with this manuscript.

\section*{Data Availability Statement}

No data are generated or analysed during this study.

\end{document}